\documentstyle[aps,multicol,prl,epsf]{revtex}
\def\sig{{\mbox{\boldmath{$\sigma$}}}}
\def\CJ{{\mbox{\boldmath{${\cal J}$}}}}
\def\CA{{\mbox{\boldmath{${\cal A}$}}}}
\newcommand{\nil}{\hspace*{0em}}
\begin{document}
\draft
\title{Orbital ac spin-Hall effect in the hopping regime }
\author{O. Entin-Wohlman,$^{1,2}$ A. Aharony,$^{1,2}$ Y. M.
Galperin,$^{3,2}$  V. I. Kozub,$^{4,2}$ and V. Vinokur$^{2}$}
\date{\today}
\address{$^1$Department of Physics,
Ben Gurion University, Beer Sheva 84105, Israel,\\ and School of
Physics and Astronomy, Tel Aviv University, Tel Aviv 69978,
Israel}
\address
{$^{2}$Argonne National Laboratory, 9700 S. Cass Ave., Argonne,
60439 IL, USA}
\address{$^{3}$Department of Physics,
University of Oslo, PO Box 1048 Blindern, 0316 Oslo, Norway}
\address{$^{4}$ A. F. Ioffe Physico-Technical Institute of the
Russian Academy of Sciences, 194021  St. Petersburg, Russia}

\maketitle
\begin{abstract}

The Rashba and Dresselhaus  spin-orbit interactions are both shown
to yield the low temperature spin-Hall effect for strongly
localized electrons coupled to phonons. A frequency-dependent
electric field ${\bf E}(\omega)$ generates a spin-polarization
current, normal to ${\bf E}$, due to interference of hopping
paths. At zero temperature the corresponding spin-Hall
conductivity is real and is proportional to $\omega^{2}$. At
non-zero temperatures the coupling to the phonons yields an
imaginary term proportional to $\omega$. The interference also
yields persistent spin currents at thermal equilibrium, at ${\bf
E}=0$. The contributions from the Dresselhaus and Rashba
interactions to the interference oppose each other.
\end{abstract}


\begin{multicols}{2}

{\it Introduction.} The possibility to control electron spins by
an electric field, due to  spin-orbit (SO) interactions, has
obvious potential applications.  Present-day research has focused
on the spin-Hall effect: An in-plane electric field applied on a
two-dimensional electron gas  creates an in-plane spin-current,
flowing in the perpendicular direction. In a two-dimensional
electron gas confined to the $x-y$ plane, the spin-current is the
flux of electrons with spins polarized along $z$.  The spin-Hall
conductivity, defined as the ratio between the spin-current and
the electric field, takes a universal value \cite{sinova} in a
pure, homogeneous infinite electron gas, but disappears
\cite{bert}   in the presence of arbitrarily small static
disorder. Consequently, the spin-Hall effect recently observed in
a two-dimensional hole system \cite{wunderlich} is apparently due
to the sample's edges.

Spin-orbit interactions in two-dimensional electronic systems
originate from bulk inversion asymmetry (due to the Dresselhaus
\cite{dresselhaus} term), or from the structural inversion
asymmetry of the potential confining the electrons to the plane
(the Rashba \cite{rashba} term). While many of the theoretical
studies concentrate on either term \cite{shekhter}, it has been
noted \cite{shen,sinitsyn,erlingsson} that there is, in fact, a
competition between the two terms, implying the intriguing
experimental possibility to control the direction of the
spin-polarization flow by modifying, e.g., the asymmetry of the
confining potential.

Whereas the spin-Hall effect of itinerant electrons in the
diffusive regime seems to be well understood, far less has been
done concerning its realization in strongly-localized electronic
systems coupled to a phonon bath \cite{damker}.  To lowest order,
the SO interaction appears as a 2$\times$2 phase-factor matrix
multiplying the hopping amplitudes \cite{oreg,raikh},
\begin{eqnarray}
\hat{J}_{i\ell}=J_{i\ell}e^{-i{\bf d}_{i\ell}\cdot\sig}.\label{J}
\end{eqnarray}
Here, $J_{i\ell}=J_{\ell i}$ is the overlap of two wave functions
localized at sites $i$ and $\ell$ (which can be chosen to be
real), $\sig$ is the vector of Pauli matrices and the vector ${\bf
d}_{i\ell}=-{\bf d}_{\ell i}$ is calculated below for the combined
Rashba and Dresselhaus interactions. Therefore, the spin-Hall
effect in insulators is due to {\em interference} of hopping
paths, reminiscing the origin of the ordinary Hall effect in
insulators \cite{holstein}. This interference also leads, as we
show below, to the appearance of `persistent spin-currents',
flowing at thermal equilibrium  in the absence of any external
fields, and surviving the coupling to the phonon bath -- again in
analogy with the ordinary Aharonov-Bohm persistent charge current
in the hopping regime \cite{arkady}. However, this spin-persistent
current does not cause any local spin accumulation \cite{rashba1}.
To achieve the latter accumulation, and to establish a spin-Hall
polarization, a frequency-dependent  electric field, ${\bf
E}(\omega)$, is required. Below we calculate the spin-Hall
conductivity matrix of a three-site triad (the smallest cluster of
localized electronic sites in which the SO interaction is
manifested), and show that it consists of off-diagonal matrix
elements alone, such that the spin-current density  in the plane,
${\bf j}^{\rm P}(\omega)$, always flows normally to the field,
\begin{eqnarray}
{\bf j}^{\rm P}(\omega)=\sigma^{\rm P}_{\rm Hall}(\omega)\left
[\begin{array}{cc}\ \ 0&\ 1\\ -1&\ 0\end{array}\right ]{\bf
E}(\omega).\label{jF}
\end{eqnarray}
At zero temperature and low $\omega$, we find that the spin-Hall
conductivity, $\sigma^{\rm P}_{\rm Hall}$, is real and quadratic
in $\omega$, and is robust against the self-averaging effect of
our highly-disordered system. As the temperature is increased, the
coupling to the thermal bath comes into play, yielding an
imaginary term in $\sigma^{\rm P}_{\rm Hall}$ which results from
dissipation and is linear in $\omega$.

As is the case for itinerant electrons
\cite{shen,sinitsyn,erlingsson}, the Rashba and the Dresselhaus
terms work one against the other: The interference phase induced
by the SO interaction is found to be proportional to the
difference between the two respective contributions. This opens
the possibility to reverse the sense of the equilibrium persistent
spin-current, as well as the direction of the spin polarization
induced by an external electric field, in the strongly localized
regime.

{\it Spin-orbit interaction in the hopping regime.}  The combined
Dresselhaus and Rashba SO interactions  in a two-dimensional
electron gas may be written in the form
\begin{eqnarray}
{\cal H}_{\rm SO}={\bf U}_{p}\cdot\sig ,\label{HSO}
\end{eqnarray}
where  ${\bf U}_{p}=[\alpha_{D}p_{x}+\alpha_{R}p_{y},
-(\alpha_{R}p_{x}+\alpha_{D}p_{y})]$ is a two-dimensional vector,
consisting of the in-plane  momentum components, and the SO
coefficients $\alpha_{D}$ (for the Dresselhaus term) and
$\alpha_{R}$ (for the Rashba one). To lowest order in the
coefficients $\alpha_{D,R}$,  the hopping matrix elements between
two states localized around sites $i$ and $\ell$ are given by Eq.
(\ref{J}), with
\begin{eqnarray}
{\bf d}_{i\ell}=[p_{D}R^{x}_{i\ell}
+p_{R}R^{y}_{i\ell},-(p_{R}R^{x}_{i\ell}+p_{D}R^{y}_{i\ell})],\label{dil}
\end{eqnarray}
where $p_{D,R}={\rm m}\alpha_{D,R}$ (m is the electronic mass, and
$\hbar$ is taken as 1) and ${\bf R}_{i\ell}={\bf R}_{i}-{\bf
R}_{\ell}$ is the difference of the radius-vectors of sites $i$
and $\ell$ \cite{note}.

Transport properties in the hopping regime are customarily
obtained as an expansion in the hopping amplitudes, since the
strong localization regime is characterized \cite{ambegaokar} by
$|J_{i\ell}|\ll|\epsilon^{\nil}_{i\ell}|$, where
$\epsilon^{\nil}_{i\ell}\equiv\epsilon^{\nil}_{i}-\epsilon^{\nil}_{\ell}$,
$\epsilon^{\nil}_{i}$ being the single-particle energies of the
localized states (assumed to be randomly distributed). The
single-bond conductance ($\propto {\rm
Tr}\{\hat{J}_{i\ell}\hat{J}_{\ell i}\}$, where the trace is in
spin space) is then independent of the SO coupling. Spin-orbit
interactions affect the hopping at third order in the $\hat{J}$'s
(and beyond). At third order, one needs to consider the
interference between the direct hopping path $i\rightarrow\ell$
and the indirect one, $i\rightarrow m\rightarrow\ell$. To second
order in the SO coefficients, the latter path involves
\begin{eqnarray}
\hat{J}_{\ell m}\hat{J}_{mi}\approx J_{\ell m}J_{mi}
(1&-&\frac{1}{2}|{\bf d}_{i\ell }|^2\nonumber\\
&-&i[{\bf d}_{\ell i} +{\bf d}_{\ell m}\times{\bf
d}_{mi}]\cdot\sig ).
\end{eqnarray}
At third order in the $\hat{J}$'s, the interference contribution
to the {\em conductance}  comes from traces like ${\rm
Tr}\{\hat{J}_{i\ell}\hat{J}_{\ell m}\hat{J}_{mi}\}$. However,
simple algebra shows that, to second order in $\alpha_{D,R}$, this
trace is independent of the SO terms.   In contrast, the
interference terms {\em do} produce a current of spin
polarization, which results from the SO interactions and is
quadratic in the ${\bf d}$'s. As discussed below, this
spin-polarization current requires the calculation of traces like
\begin{eqnarray}
\CJ_{i\ell m} \equiv -i{\rm Tr}\{\sig\hat{J}_{i\ell}\hat{J}_{\ell
m}\hat{J}_{mi}\}.\label{TrJ}
\end{eqnarray}
The cornerstone of all our results is the observation that
\begin{eqnarray}
\CJ_{i\ell m}= 4J_{i\ell}J_{\ell m}J_{mi}(p_{D}^{2}-p_{R}^{2})
{\CA},\label{JJJ}
\end{eqnarray}
where the vector ${\CA}={\bf R}_{mi}\times{\bf R}_{i\ell}/2$ is
along the $z$ direction, and $|{\CA}|$ is equal to the area
spanned by the three sites, $i$, $\ell$ and $m$, in the $x-y$
plane. It is also seen that the two SO interactions considered
here have opposite effects on the interference phase.

 {\it Spin polarization and spin-polarization current
density in the hopping regime.} In our calculations we employ the
usual electron-phonon Hamiltonian pertaining to electrons in the
hopping regime, ${\cal H}={\cal H}_{0}+{\cal V}$, in which
\begin{eqnarray}
{\cal
H}_{0}=\sum_{n\sigma}\epsilon^{\nil}_{n}n^{\nil}_{n\sigma}+\sum_{\bf
q}\omega^{\nil}_{q}b^{\dagger}_{\bf q}b^{\nil}_{\bf q},\ \
n^{\nil}_{n\sigma}=c^{\dagger}_{n\sigma}c^{\nil}_{n\sigma},
\end{eqnarray}
where $c^{\dagger}_{n\sigma}$ ($c^{\nil}_{n\sigma}$) creates
(destroys) an electron of spin index $\sigma$ on the state
localized at site $n$, $\omega^{\nil}_{q}$ is the phonon
frequency, and $b^{\dagger}_{\bf q}$, $b^{\nil}_{\bf q}$ are the
phonon operators. The Hamiltonian ${\cal V}$ includes the hopping,
the coupling to the phonons \cite{arkady}, and the external
electric field,
\begin{eqnarray}
{\cal V}=\sum_{nn'}\sum_{\sigma\sigma
'}\hat{J}_{nn'}^{\sigma\sigma '}Q_{nn'}^{\nil}
e^{i\Phi^{\nil}_{nn'}(t)}c^{\dagger}_{n\sigma}c^{\nil}_{n'\sigma
'},
\end{eqnarray}
where the spin-dependent hopping is defined in Eq. (\ref{J}),
\begin{eqnarray}
Q_{nn'}^{\nil}=\exp [\sum_{\bf q}\frac{v_{\bf
q}^{nn'}}{\omega^{\nil}_{q}}(b^{\dagger}_{-{\bf q}}-b^{\nil}_{\bf
q})]
\end{eqnarray}
yields the effect of an on-site electron-phonon coupling ($v_{\bf
q}^{nn'}=v^{n}_{\bf q}-v^{n'}_{\bf q},$ with $v^{n}_{\bf q}$
representing the electron-phonon coupling on site $n$) on the
hopping, and $\Phi^{\nil}_{nn'}$ is due to an in-plane external
electric field, with $i\omega\Phi^{\nil}_{nn'}(\omega )=e{\bf
E}\cdot{\bf R}_{nn'}$ \cite{yuri}.

The spin polarization operator at site $n$ is given by $P_{n}={\rm
Tr}\{\sigma^{z}n^{\nil}_{n\sigma}\}$. The spin-polarization
current density, ${\bf j}^{\rm P}$, is then defined in analogy
with the current density due to an electrical dipole-moment
\cite{holstein},
\begin{eqnarray}
{\bf j}^{\rm P}=
%\frac{1}{|S|}
(1/S)\sum_{n}\langle
%\frac{dP_{i}}{dt}
dP_{n}/dt\rangle{\bf R}_{n},\label{j}
\end{eqnarray}
where $S$ denotes the area of the system. Below we concentrate on
${\bf j}^{\rm P}$ of  a single triad, and therefore this area is
replaced by $|\CA |$. The average, $\langle\ldots\rangle$, is
calculated with the Hamiltonian ${\cal H}$ assuming that each site
is in contact with a grand-canonical electron reservoir
\cite{holstein,arkady}. The temporal derivative $dP_{n}/dt$ is
found by calculating $i[{\cal H},n_{n\sigma}^{\nil}]$ up to
first-order in the electric field. This yields for the  average
(carried out with the Hamiltonian ${\cal H}$)
\begin{eqnarray}
\langle dP_{n}/dt \rangle =&-& \sum_{n'}\sum_{\sigma\sigma
'}\sigma^{z}_{\sigma\sigma }\langle I^{\sigma\sigma '
}_{nn'}-I^{\sigma '\sigma}_{n'n}\nonumber\\
&-& i\Phi_{nn'}(t)(I^{\sigma\sigma '}_{nn'}+I^{\sigma '\sigma
}_{n'n})\rangle ,\label{PT}
\end{eqnarray}
with
\begin{eqnarray}
I^{\sigma\sigma '}_{nn'}&=&i Q^{\nil}_{nn' }
c^{\dagger}_{n\sigma}\hat{J}^{\sigma\sigma '}_{nn' }c_{n'\sigma
'}^{\nil}.\label{IP}
\end{eqnarray}
We calculate $\langle dP_{n}/dt \rangle$ to third-order in the
hopping, and up to first-order in the electric field. Detailed
calculations \cite{com1} show that $\langle I^{\sigma\sigma '
}_{nn'}+I^{\sigma '\sigma}_{n'n}\rangle$, to third-order in the
hopping, vanishes.  The entire contributions to the polarization
rate comes from $\langle I^{\sigma\sigma '}_{nn'}-I^{\sigma
'\sigma}_{n'n}\rangle$ in (\ref{PT}), which will be found up to
first-order in the electric field.

{\it Persistent spin currents.} The thermal average $\langle
I^{\sigma \sigma '}_{nn'}-I^{\sigma '\sigma}_{n'n}\rangle$ is
non-zero {\em even} when the electric field is absent. This is a
consequence of the trace Eq. (\ref{TrJ}), which includes $\sig$:
Since the SO interaction conserves time-reversal symmetry, it
cannot produce persistent charge currents, but it {\em does} lead
to persistent spin-current. Indeed, ignoring the electron-phonon
coupling, and using third-order perturbation theory in the hopping
matrix elements, the contribution of the bond $i-\ell$ in our
triad (at ${\bf E}=0$) becomes
\begin{eqnarray}
\sum_{\sigma\sigma '}\sigma^{z}_{\sigma\sigma}\langle I^{\sigma '
\sigma }_{\ell i}-I^{\sigma \sigma '}_{i\ell } \rangle |_{\rm
eq}=2{\cal J}_{i\ell m}^{z}\sum_{\rm
per}\frac{f_{i}}{\epsilon^{\nil}_{i\ell}\epsilon^{\nil}_{im}},\label{pers}
\end{eqnarray}
where $f_{i}$ is the Fermi function of the site $i$ occupation,
and $\sum_{\rm per}$ stands for the sum over the three
permutations $i,\ell,m\rightarrow\ell ,m, i\rightarrow m,i,\ell$.

Hence, there is a persistent spin-current flowing around the triad
at equilibrium, even at zero temperature, with its sense being
determined by the relative locations of the site energies compared
to the Fermi level. This finding is in a complete analogy with the
persistent charge current flowing in response to an Aharonov-Bohm
phase \cite{arkady}. The analogy continues when the coupling to
the phonon bath is switched-on: That coupling induces a
Debye-Waller factor multiplying the result (\ref{pers}), {\it and
also} a `counter' spin-persistent current, which flows in the
opposite direction to the current (\ref{pers}), and vanishes at
zero temperature (for details, see Ref. \onlinecite{arkady}). This
equilibrium persistent current does not lead to spin accumulation:
The current flowing from $i$ to $\ell$ is minus the one flowing
from $i$ to $m$. [When $\ell$ is interchanged with $m$, the sign
of $\CA$ is reversed, see Eq. (\ref{JJJ}).] Consequently $\langle
dP_{i}/dt\rangle |_{\rm eq}$ vanishes (an analogous calculation
shows that at equilibrium, $\langle P_{i}\rangle |_{\rm eq}$
vanishes as well). Equilibrium spin-currents which do not lead to
spin accumulation exist also in conductors lacking inversion
symmetry \cite{rashba1}. The question whether they are amenable to
an experimental detection (like the charge persistent currents) in
small coherent mesoscopic structures is left open.

 {\it Spin-Hall effect in the hopping regime.} When an
in-plane electric field is applied, the polarization rate $\langle
dP_{i}/dt\rangle$ is non-zero. We first consider it for our
three-site cluster  in the absence of the coupling to the phonon
bath, employing linear response theory (with respect to the field)
and third-order perturbation theory (with respect to the hopping
amplitudes). The result is
\begin{eqnarray}
&&\langle dP_{i}/dt\rangle (\omega
)\nonumber\\
&\equiv&\sum_{\sigma\sigma '}\sigma^{z}_{\sigma\sigma} \langle
I^{\sigma '\sigma} _{\ell i}(\omega )-I^{\sigma\sigma '}_{i\ell
}(\omega ) + I^{\sigma '\sigma}_{mi}(\omega )- I^{\sigma\sigma
'}_{im}(\omega )\rangle \nonumber\\
& =&\omega^{2}\Phi_{\ell m}(\omega )\frac{2{\cal J}^{z}_{i\ell
m}}{\epsilon^{\nil}_{i\ell}\epsilon^{\nil}_{\ell
m}\epsilon^{\nil}_{mi}}\sum_{\rm
per}\frac{f_{i}-f_{\ell}}{\omega^{2}-\epsilon_{i\ell}^{2}}.\label{PDOT0}
\end{eqnarray}
Namely, the average polarization rate at site $i$ is driven by the
potential difference across the bond $\ell -m$. This implies that
inserting Eq. (\ref{PDOT0}) into Eq. (\ref{j}) produces Eq.
(\ref{jF}), with the spin-Hall conductivity, $\sigma^{\rm P}_{\rm
Hall}$, given by
\begin{eqnarray}
\sigma^{\rm P}_{\rm Hall}(\omega
)=16e\omega^{2}(p_{R}^{2}-p_{D}^{2})|\CA|\Gamma^{\rm e}_{i\ell
m}(\omega ),\label{SIG0}
\end{eqnarray}
where
\begin{eqnarray}
\Gamma^{\rm e}_{i\ell m}(\omega )=\frac{J_{i\ell}J_{\ell
m}J_{mi}}{\epsilon^{\nil}_{i\ell}\epsilon^{\nil}_{\ell
m}\epsilon^{\nil}_{mi}}\sum_{\rm
per}\frac{f_{i}-f_{\ell}}{\epsilon^{2}_{i\ell}-\omega^{2}}.\label{Gamma0}
\end{eqnarray}
In deriving this result, we have discarded resonance transitions
in which  the frequency $\omega$ compensates the site energy
differences. Otherwise, the sum in Eq. (\ref{Gamma0}) would have
been augmented by  terms of the type $i(f_{i}-f_{\ell})\delta
(\epsilon^{\nil}_{i\ell}\pm \omega )/|\omega |$. We will find
below that the electron-phonon coupling gives rise to an imaginary
part in the spin-Hall conductivity, that is of the same order in
$\omega$, and which originates from energy-conserving
delta-functions of the type $\delta
(\epsilon_{i\ell}^{\nil}\pm\omega_{q}\pm\omega)$. Since the latter
are more likely to be satisfied for the randomly-distributed site
energies, the present discussion is confined to the result
(\ref{SIG0}).

The result (\ref{jF}) demonstrates in a nutshell the origin of the
present-day interest in the spin-Hall effect. While an electric
field along, say the $x$ direction, will drive a charge current in
the same direction, the spin-current density in this case will be
along the $y$ direction, leading to a `separation' of spin and
charge.

The spin-Hall conductivity (\ref{SIG0}) vanishes at small
frequencies and tends to zero as $\omega ^{-2}$ at very high ones
[since when $\epsilon^{2}_{i\ell}\ll\omega^{2}$ the sum in Eq.
(\ref{Gamma0}) vanishes]. Thus, a dc electric field is incapable
of producing the spin-Hall effect. This is in accordance with the
behavior found in itinerant electron systems \cite{bert}. This
conductivity, which depends upon the temperature through the Fermi
functions, remains finite at zero temperature when two of the
three site energies are below or above the Fermi energy. Then, at
frequencies smaller than the typical site energy differences, and
assuming that all the hopping matrix elements $J_{i\ell}$ have the
same sign \cite{raikh}, $\Gamma^{\rm e}_{i\ell m}$ takes a
definite sign, independent of the locations of the site energies
relative to the Fermi level. Consequently, the self-averaging
effect of the macroscopic hopping system, which involves many
triads will not wash out $\sigma^{\rm P}_{\rm Hall}$.

The effect of the coupling to the phonons on the above results may
be divided into three. Firstly, there is the overall Debye-Waller
exponent, due to loss of coherence. This factor exists even at
zero temperature. (For brevity, it will not be presented below.)
Secondly, virtual electron-phonon processes contribute additional
terms to $\Gamma^{\rm e}_{i\ell m}$, which are proportional to the
electron-phonon coupling and involve complicated combinations of
site Fermi functions and Bose occupation numbers of the phonons.
We do not present these terms since in the weak electron-phonon
coupling they are smaller than the purely electronic ones given in
Eq. (\ref{Gamma0}). Thirdly and most importantly, real
(energy-conserving) electron-phonon transitions give rise to an
imaginary part in the Hall-conductivity, which now becomes
\begin{eqnarray}
\sigma^{\rm P}_{\rm Hall}(\omega
)=16e(p_{R}^{2}-p_{D}^{2})|\CA|(\omega^{2}\Gamma_{i\ell m}^{\rm
e}(\omega )-i\Gamma^{\em ep}_{i\ell m}(\omega ))\label{HALLP}
\end{eqnarray}
with
\begin{eqnarray}
\Gamma^{\rm ep}_{i\ell m}(\omega )=\frac{{\rm sinh}(\beta\omega
)}{2\beta}\sum_{\rm per}(\frac{1}{\epsilon^{2}_{\ell
m}}+\frac{1}{\epsilon^{2}_{mi}})\frac{G_{i\ell}}{e^{2}}\frac{J_{\ell
m}J_{mi}}{J_{i\ell}},\label{WG}
\end{eqnarray}
where $G_{i\ell} \propto \exp[-\beta
(|\epsilon^{\nil}_{i\ell}|+|\epsilon^{\nil}_{i}|
+|\epsilon_{\ell}^{\nil}|)/2]$ is the usual temperature-dependent
hopping conductance of the $i-\ell$ bond, and $\beta$ is the
inverse temperature. The result (\ref{WG}) is derived in the small
$\omega$ limit, assuming the frequency to be smaller than the site
energy differences. Hence, $\Gamma^{\rm ep}$ vanishes at zero
temperature.

To obtain the temperature dependence of the Hall-conductivity
(\ref{HALLP}) at low frequencies, we consider the situation in
which the leading electrical conductance of our triad takes place
along the bond $i-\ell$, and site $m$ provides the interference
path necessary for the spin-Hall effect. Thus we imagine
$\epsilon_{i}^{\nil}$ and $\epsilon_{\ell}^{\nil}$ to be below and
above the Fermi level, but close to it, while
$\epsilon_{m}^{\nil}$ lies far away from the Fermi energy. In that
case $\Gamma_{i\ell m}^{\rm e}\sim J_{i\ell}J_{\ell
m}J_{mi}/|\epsilon^{3}_{i\ell}|\epsilon_{m}^{2}$, while
$\Gamma^{\rm ep}_{i\ell m}\sim\Gamma^{\rm e}_{i\ell
m}\gamma_{i\ell}{\rm sinh}(\beta\omega
)|\epsilon^{\nil}_{i\ell}|\exp (-\beta
|\epsilon_{i\ell}^{\nil}|)$, where $\gamma_{i\ell}=|v|^{2}{\cal
N}(|\epsilon^{\nil}_{i\ell}|)$ is the density of phonon states at
$|\epsilon^{\nil}_{i\ell}|$ multiplied by the electron-phonon
matrix element squared. It then turns out that one may define a
characteristic frequency of the system \cite{yuri},
\begin{eqnarray}
\nu_{i\ell}(T)=\gamma_{i\ell}\beta|\epsilon_{i\ell}^{\nil}|e^{-\beta
|\epsilon^{\nil}_{i\ell}|},
\end{eqnarray}
which vanishes at zero temperature. For $\beta\omega <1$ the
spin-Hall conductivity is then
\begin{eqnarray}
\sigma^{\rm P}_{\rm Hall}(\omega
)\propto\omega[\omega-i\nu_{i\ell}(T)].
\end{eqnarray}
At very low temperature, this conductivity is proportional to
$\omega^{2}$; As the temperature is  increased, real
electron-phonon processes give  rise to an imaginary part due to
dissipation,  which is linear in $\omega$.

In conclusion, we have shown that the competition between the
Rashba and the Dresselhaus SO interactions opens the possibility
to control the magnitude and the phase (compared to the driving ac
field) of the spin-Hall current for localized electrons in the
hopping regime.

\vspace{0.5cm} \noindent  We acknowledge helpful discussions with
A. M. Finkel'stein and Y. Imry. This project was carried out in a
center of excellence supported by the ISF under grant No. 1566/04.
Work at Argonne is supported by  the U.S. Department of Energy
under contract W-31-109-Eng-38.

\end{multicols}


\begin{references}
\bibitem{note}
Equation (\ref{dil}) was obtained in Ref. \onlinecite{raikh} for
the Rashba case.
\bibitem{com1}
Because of the spin summations, it turns out that this average,
%$\langle \Re I^{\rm P}_{i\ell}\rangle$
in a three-site cluster, requires the imaginary part of $\langle
Q_{nn'}c^{\dagger}_{n\sigma}c^{\nil}_{n'\sigma '}\rangle$, which
is shown in Ref. \onlinecite{arkady} to vanish.
\end{references}
\end{document}